\documentclass[useAMS,usenatbib]{mn2e}

\usepackage{graphics,epsfig}
\usepackage{amsmath}
\usepackage{amssymb}
\usepackage{graphicx}

\newcommand{\kms}{km~s$^\mathrm{-1}$}
\newcommand{\sunn}{$_{\odot}$}
\newcounter{qub}
\newcommand{\NHI}{N_{\rm HI}}

\DeclareRobustCommand{\ion}[2]{%
\relax\ifmmode
\ifx\testbx\f@series
{\mathbf{#1\,\mathsc{#2}}}\else
{\mathrm{#1\,\mathsc{#2}}}\fi
\else\textup{#1\,{\mdseries\textsc{#2}}}%
\fi}

\title{\ion{H}{i} in XMD Galaxies III. GMRT observations of BCG HS~0822+3542}

\author[Jayaram N. Chengalur, S. A. Pustilnik, J.-M. Martin, A.Y. Kniazev]
{Jayaram N. Chengalur$^{1,2}$\thanks{chengalur@ncra.tifr.res.in}
 S.A. Pustilnik$^{3}$\thanks{sap@sao.ru}
 J.-M. Martin$^{4}$\thanks{jean-michel.martin@obspm.fr}
 A.Y. Kniazev$^{5,3}$\thanks{akniazev@saao.ac.za}\\
$^1$ ATNF/CSIRO, P. O. Box 76, Epping NSW 1710, Australia \\
$^2$ NCRA (TIFR), Pune University Campus, Postbag 3, Ganeshkhind,
     Pune 411007, India \\
$^3$ Special Astrophysical Observatory RAS, Nizhnij Arkhyz,
     Karachai-Circassia,  369167 Russia \\
$^4$ Observatoire de Paris, 5, place J. Janssen, 92195 Meudon Cedex, France \\
$^5$ South African Astronomical Observatory, Cape Town, South Africa
}

\begin{document}

\label{firstpage}

\date{Accepted 2006 . Received 2006 }

\pagerange{\pageref{firstpage}--\pageref{lastpage}} \pubyear{2005}

\maketitle

\begin{abstract}

 We present the results of Giant Metrewave Radio Telescope
 \ion{H}{i} 21-cm line observations of the eXtremely Metal Deficient
(XMD) blue compact galaxy (BCG) HS~0822+3542. HS~0822+3542 is 
the smallest known XMD galaxy; from HST imaging it has been suggested 
that it actually  consists of two still smaller ($\sim$100~pc sized)
ultra-compact dwarfs that are in the process of merging. The
brighter of these two putative ultra compact dwarfs has an
ocular appearance, similar to that seen in galaxies that have
suffered a penetrating encounter with a smaller companion.
From our \ion{H}{i} imaging we find that the gas distribution
and kinematics in this object are similar to that of other low
mass galaxies, albeit with some evidence for tidal disturbance. 
On the other hand, the \ion{H}{i} emission has an angular 
size $\sim$25 times  larger than that of the putative 
ultra-compact dwarfs. The optical emission is also
offset from the centre of the \ion{H}{i} emission. 
HS~0822+3542 is located in the nearby Lynx-Cancer void, but 
has a nearby companion LSB dwarf galaxy  SAO~0822+3545. 
In light of all this we also consider a scenario
where the optical emission from HS~0822+3542 comes not from
two merging ultra-compact dwarfs but from multiple star forming regions
in a tidally disturbed galaxy. In this model, the ocular
appearance of the brighter star forming region
could be the result of triggered star formation.

\end{abstract}

\section{Introduction}
\label{sec:intro}

     The metallicity of the ISM of external galaxies is often quantified
in terms of the oxygen abundance,  which, in turn, is generally expressed
in terms of the quantity 12+$\log$(O/H). A very small fraction of gas-rich
dwarfs have 12+$\log$(O/H) $\lesssim$7.65, i.e. an oxygen abundance of 
less than $\sim$1/10 solar. These dwarfs, (the majority of which are 
also members  of Blue Compact Galaxy (BCG) class), are termed eXtremely
Metal Deficient (XMD) galaxies (see \cite{kunth00} for a review).
Low oxygen abundances are naturally expected for galaxies that have only 
recently experienced their first episode of star formation.
For example, metallicities of $\lesssim$1/10 solar are typical of high
redshift Damped Lyman-$\alpha$ systems (e.g., \citealp{pettini97}),
which are widely regarded as proto-galactic objects. It is now
well established that the majority of (but not all) well studied XMD galaxies 
have a substantial mass fraction of their stellar mass in old stars 
(see e.g. a brief review in \citealp{pustilnik06}) and are hence not
forming stars for the first time. Nonetheless, their very low 
ISM metallicity still makes them the best local analogs 
of young galaxies at high redshift. Detailed studies of 
XMD galaxies should hence give insight into galaxy evolution in 
the early universe.

  While the {\it fraction} of XMD BCGs in the Local Universe is very small
($\lesssim$2\% of all known BCGs, e.g., \citealp{pustilnik05}), 
the {\it number} of known XMD galaxies has seen a dramatic increase 
in the recent past, via new large surveys for emission line galaxies 
(e.g. HSS, KISS and SDSS), and $\sim$60 such galaxies are now known. 
However, detailed optical and radio observations are currently 
available only for a very small fraction of them.  As less evolved 
galaxies, XMDs are  expected  to be very gas-rich, and indeed for 
some objects (e.g., SBS 0335$-$052;  \citealp{pustilnik01,pustilnik04}), 
the gas makes up $\sim 99$\% of the total baryonic mass. 
\ion{H}{i} observations
are hence critical in understanding the nature of these rare 
galaxies. For some of XMD BCGs from the Second Byurakan Survey (SBS) 
the integrated \ion{H}{i} parameters are available in \cite{thuan99}.
More recently, \cite{pustilnik06} (Paper I) conducted single dish
HI observations for 22 XMD galaxies. Follow-up aperture synthesis 
\ion{H}{i} observations of these galaxies are now in progress 
(e.g. BCG SBS 1129+576 \cite{ekta06}, (Paper II)).

In this paper we present Giant Metrewave Radio Telescope (GMRT) 
\ion{H}{i}~21cm observations of one of the nearest 
known XMD BCGs, HS 0822+3542 (12 + log(O/H) = 7.35;
\citealp{kniazev00}). \cite{pustilnik03} found that this galaxy 
has a companion at a projected separation of 3.8\arcmin\ 
(i.e. $\sim$11.4 kpc at the distance to the pair\footnote{
We assume a distance of 11~Mpc. At this distance 1\arcsec\ 
corresponds to a linear separation of $\sim$53~pc.}) --  
the Low Surface Brightness (LSB) dwarf galaxy SAO 0822+3545. The 
separation between HS~0822+3542 and SAO~0822+3545 is much 
smaller than that of the primary beam of the GMRT antennas, and our 
observations are sensitive to \ion{H}{i} emission from both of them. 
In Sec.~\ref{sec:obs} we describe the GMRT observations and data
reduction. Results are presented in Sec.~\ref{sec:res} and discussed
in Section~\ref{sec:dis}. 

\section{Observations and reduction}
\label{sec:obs}

\begin{table}
\caption{Parameters of the GMRT observations}
\label{tab:obs}
\vskip 0.1in
\begin{tabular}{ll}
\hline
Parameters& Value \\
\hline
\hline
RA(2000)                & $08^h 25^m 55.4^s$\\
Declination(2000)       & $35^{\circ} 32' 32''$\\
Date of observations    & 18 \& 30 June 2001\\
Time on source          & 7 hrs\\
Total bandwidth         & 2.0 MHz\\
Number of channels      & 128\\
Channel separation            & 3.3 km sec$^{-1}$\\
FWHM of synthesized beam      & $42''\times37'', 27''\times23''$,\\
			      & $17''\times15'', 4''\times3''$ \\
RMS (at 6.6 \kms\ resolution) & 1.3~mJy, 1.1~mJy\\
                              & 1.0~mJy, 0.7~mJy\\
\hline
\end{tabular}
\end{table}

    GMRT (\citealp{swarup91}) observations of HS~0822+3542 were conducted 
on 18th and 30th June, 2001, when the telescope was in its commissioning 
phase. The setup for the  observations is given  in Table~\ref{tab:obs}. 
Absolute flux calibration  and bandpass calibration were done using scans
on the standard calibrators 3C48 and 3C286, one of which was observed 
at the start and end of each observing run. Phase calibration was done 
using the VLA calibrator source 0741+312 which was observed once 
every 45 minutes. The data were reduced using standard tasks in classic AIPS. 
For each run, bad visibility points were edited out, after which the 
data were calibrated. Calibrated data for both runs were combined using DBCON. 
The GMRT does not do online Doppler tracking -- the differential Doppler 
shift between the two runs was hence corrected for using the AIPS task 
CVEL before combining the two observing runs. The GMRT has a hybrid 
configuration and a single GMRT observation can be used to make maps
at a range of angular scales. Data cubes were therefore made at 
resolutions ranging from $\sim$40$^{''}$ to $\sim$3$^{''}$.
Continuum was subtracted using the AIPS task IMLIN, and the data cubes
were cleaned using APCLN. Finally moment images were made using the
AIPS task MOMNT.

    The \ion{H}{i} emission from the galaxy pair
spans 24 channels of the spectral cube. A continuum image was made using the
average of remaining line free channels. The only detected emission was from
a faint ($\sim$7~mJy) source at 
08$^h$25$^m$59.3$^s$+35$^{\degr}$31$^{'}$12$^{''}$, i.e.
1\farcm2 southeast of the \ion{H}{i} emission seen from HS~0822+3542.
Given that the location of the continuum emission is outside the \ion{H}{i}
disk, it appears likely that it is a background source, not associated 
HS~0822+3542. This is consistent with the NED entries for this position
(i.e. a FIRST source with $f_{\nu}$(1.4~GHz) $\sim$4~mJy, and a 2MASS 
galaxy).

\section{Results}
\label{sec:res}

\subsection{Morphology and density distribution}
\label{morphology}

\begin{figure}
\begin{center}\includegraphics[width=7cm]{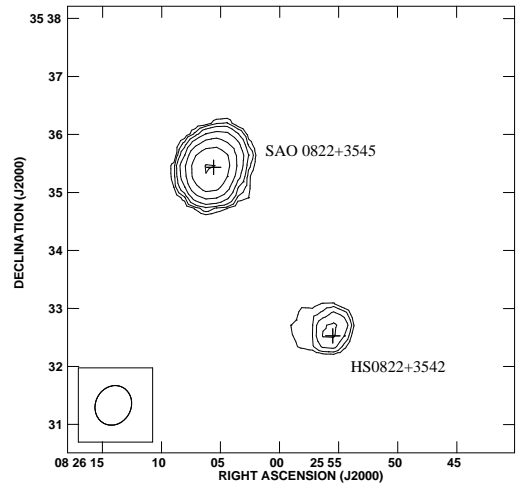}\end{center}
\caption{
  GMRT $42^{''}\times37^{''}$  resolution integrated \ion{H}{i}
  emission map from HS~0822+3542 and SAO~0822+3545. The contour 
  levels are at $2.5$ (i.e $\sim 4\sigma$), $4.4, 7.7, 13.3, 21.3$, 
  and $40.2 \times 10^{19}$  atoms~cm$^{-2}$. The crosses mark 
  the centres of the optical  galaxies.
}
\label{fig:uv5m0}
\end{figure}

\begin{figure}
\begin{center}\includegraphics[width=8cm]{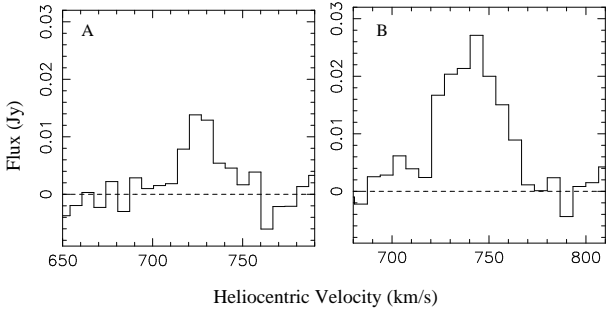}\end{center}
\caption{Synthetic single dish \ion{H}{i} profiles for  HS~0822+3542 (panel A)
   and SAO~0822+3545 (panel B). The channel separation is $6.6$~\kms. }
\label{fig:uv10spc}
\end{figure}

The integrated \ion{H}{i} emission from HS~0822+3542 and SAO~0822+3545 
at an angular resolution of $42^{''}\times37^{''}$ is
shown in Fig.~\ref{fig:uv5m0}. Synthetic single dish spectra are shown
in Fig.~\ref{fig:uv10spc}. There is a hint of tidal distortion seen
in the lowest contour of HS~0822+3542. The parameters derived from the
synthetic profiles are summarised in Tab.~\ref{tab:param}. The bulk of 
the \ion{H}{i} flux from this pair comes from the companion LSB galaxy
SAO~0822+3545. If the emission from the two galaxies is added together, 
the resulting profile (not shown) is a reasonable match to that obtained
at the NRT by \cite{kniazev00}, where the beam covered both galaxies. 
As expected, the total flux obtained by the NRT is somewhat lower 
than that obtained by the GMRT. This is because the brighter galaxy 
in the pair  was close  to the half-power point of the NRT beam.

\begin{figure}
\begin{center}\includegraphics[width=9cm]{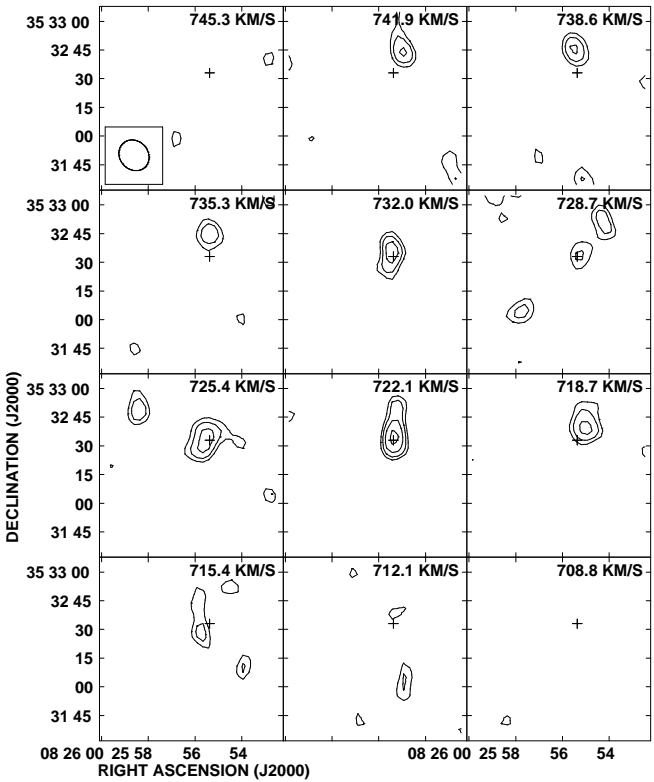}\end{center}
\caption{ 
  Channel maps of the \ion{H}{i} emission from HS~0822+3542 at
  a resolution of $17^{''}\times15${''}. The contours are at 
  3.4, 4.4, 5.6 and 7.2 mJy/Bm. The cross marks the position
  of the optical galaxy}
\label{fig:chan}
\end{figure}

In Fig.~\ref{fig:chan}  we show channel maps of the emission from
HS~0822+3542 at a resolution of $17^{''}\times15{''}$. The signal-to-noise
ratio is marginal, but it can still be seen that the
centroid of the emission shifts from north to south as one moves 
towards lower heliocentric velocities. 

\begin{figure*}
\begin{center}\includegraphics[width=12cm]{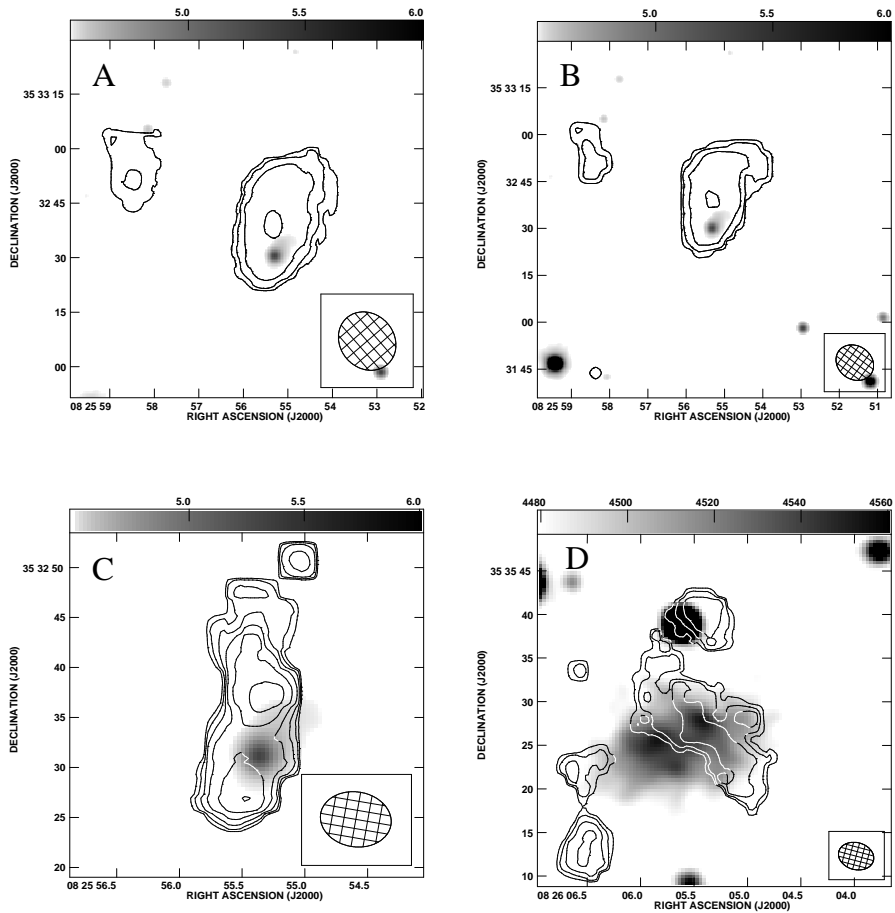}\end{center}
\caption{ Overlay of the SDSS $r$-band optical emission (grey scales)
   on the integrated \ion{H}{i} emission (contours) from HS~0822+3542 and
   SAO~0822+3545.
     [A] HS~0822+3545, resolution $17^{''}\times15{''}$. The contours 
         are at 0.52, 1.2, 2.6 and 5.7 $\times 10^{20}$ atoms~cm$^{-2}$. 
     [B] HS~0822+3545, resolution $13^{''}\times11{''}$. The contours 
         are at 0.23, 0.75, 2.4 and 7.8 $\times 10^{20}$ atoms~cm$^{-2}$. 
     [C] HS~0822+3545, resolution $7^{''}\times5.5{''}$. The contours are 
         at 2.3, 3.3, 4.6, 6.6, 9.3 and 13.2 $\times 10^{20}$ atoms~cm$^{-2}$. 
     [D] SAO~0822+3545, resolution $4^{''}\times3.1{''}$. The contours are 
         at 0.74, 1.3, 2.2 and 3.7 $\times 10^{21}$ atoms~cm$^{-2}$.  
         Note that the optical emission seen coincident with the northern
         most HI emission is an unrelated foreground star.
}
\label{fig:m0ovhr}
\end{figure*}

In Fig.~\ref{fig:m0ovhr} we show overlays of the SDSS $r$-band (\citealp{DR3})
optical emission and the \ion{H}{i} emission from HS~0822+3542 
and SAO~0822+3545. The \ion{H}{i} extension to the north-east of 
HS~0822+3545 seen in Fig.~\ref{fig:uv5m0} can also be seen
in  Fig.~\ref{fig:m0ovhr}[A] \& [B], although at these resolutions
the emission does not connect to the main body. This could be
a consequence of both the marginal signal-to-noise ratio as well
as the lower sensitivity to diffuse emission in these maps.
One can also see a tidal tail like extension from the northern
end of HS~0822+3542 (most clearly in panel [B]). The \ion{H}{i} 
emission (apart from these features) is fairly symmetric
about its peak. The \ion{H}{i} emission is unusually extended compared
to the optical emission, but even more unusual is the large offset 
between the optical emission and the centre of the \ion{H}{i} emission.
While the \ion{H}{i} distribution seems in general uncorrelated with
the optical emission, one can see in Fig.~\ref{fig:m0ovhr}[C] a 
kink in the \ion{H}{i} contours 
aligned with the diffuse plume of optical emission extending to 
the north-west.

\begin{figure}
\begin{center}\includegraphics[width=8cm]{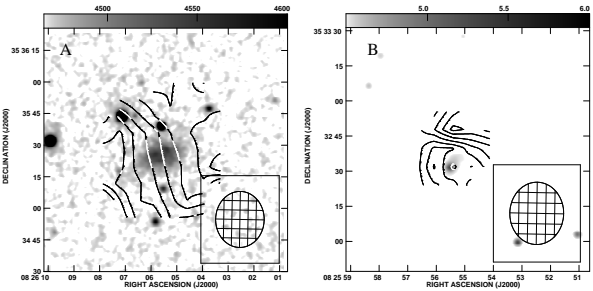}\end{center}
\caption{
   [A]~The velocity field of SAO~0822+3545 as derived from the
       27$^{''}\times 23^{''}$ resolution  data cube. The velocity 
       contours are at 732, 735, 738, 741, 744, 747, 750 753 and 
       756 ~\kms\ (heliocentric). The grey scale is the SDSS $r$-band
       image.
   [B] The velocity field of HS~0822+3542 as derived from the
       27$^{''}\times 23^{''}$ resolution  data cube. The velocity 
       contours are at 722, 723, 724, 725, 726, 727 and 728~\kms\
       (heliocentric). The grey scale is the SDSS $r$-band image.
}
\label{fig:m1ov}
\end{figure}

   The velocity fields for SAO~0822+3545  and HS~0822+3542 (as derived 
from the first moment of the of the 27$^{''}\times 23^{''}$ resolution 
data cube) are shown in Fig.~\ref{fig:m1ov}. We note that the spatial
resolution is marginal (particularly so for HS~0822+3542); the inferences 
we make below should hence be regarded as tentative.  

    For SAO~0822+3545, the velocity field appears regular, with an 
indication of a warp in the outer parts, i.e. starting from the edge 
of the optical disk. From an inspection of the  velocity field, one 
can crudely estimate the inclination corrected (where the inclination
is computed from the optical image, see Table~\ref{tab:param}) rotation 
velocity to 
be $\sim 14$~\kms. For HS~0822+3542, a north-south velocity gradient,
consistent with that seen in the channel maps is seen in the velocity 
field. If this is indicative of rotation, the implied rotational 
velocity is extremely 
small, viz. $\sim$4~\kms (i.e. smaller than the velocity dispersion,
see below).  

    The velocity dispersion, $\sigma$,  in galaxies of comparable 
\ion{H}{i} mass to SAO~0822+3545 and HS~0822+3542 is 
\mbox{$\sigma \sim$7--9~\kms}\ (see, e.g., \citealp{begum06}), 
i.e., comparable to the peak rotation speed in the galaxies.
Clearly, both rotation as well as random velocities are important in
supporting the \ion{H}{i} disk against collapse. Determining the 
dynamical mass from the observed velocity field hence requires 
correction for the support provided by the pressure gradient in 
the disk. Computing this requires one to have a good estimate of 
the density gradient across the disk (see, e.g., \cite{begum03}; 
\citealp{begum04}). The modest signal-to-noise ratio and spatial 
resolution of our data would result in substantial systematic 
uncertainties, were we to try this approach. We instead compute 
an indicative dynamical mass following \cite{lister92}.
In detail, we assume that  $W_{50} \sim 2(V_{rot}\sin(i)+1.18\sigma)$ 
and M$_{\rm dyn} = 3.3 \times 10^{4} a_H D (V_{\rm rot}^2 + 3\sigma^{2})$,
where $D$ is the distance in Mpc and $a_H$ is the diameter in arcmin.
For SAO~0822+3545 taking $a_H \sim$1.5, $\sigma \sim$7,
and the inclination listed in Tab~\ref{tab:param}, gives 
$V_{\rm rot} \sim$10~\kms\ and 
$M_{\rm dyn} \sim$1.4$\times$10$^{8}$~M$_\odot$. Assuming a similar
velocity dispersion for HS~0822+3542 and an angular diameter 
of $\sim$0.75$^{'}$ gives $V_{\rm rot} \sim$3~\kms\ and $M_{\rm dyn}
\sim$4$\times$10$^{7}$~M$_\odot$. The low ratio of $V_{\rm rot}/\sigma$
in HS~0822+3542 is typical of very low mass dwarf galaxies 
(see eg. \citealp{begum06}).

\section{Discussion}
\label{sec:dis}
   
 The main \ion{H}{i} and optical parameters of HS~0822+3542 and SAO~0822+3545
are summarised in Tab.~\ref{tab:param}. Using deconvolved Nordic
Optical Telescope images, 
Pustilnik et al. (2003) found that the bright knot of optical emission 
in HS~0822+3542 consists of two components.
Based on  HST images, \cite{corbin05} state that these two components
have a  surface brightness that is $\sim 100$ times larger than 
that of the diffuse emission extending to the north west. The
brighter of these two components (component ``A'' in the terminology
of \cite{corbin05}; see their Fig~1) has an ocular morphology, 
with a central blue star forming region surrounded by a ring of 
redder stars. The fainter component (component ``B'') is
$\sim$5 times less luminous and more irregular in appearance. Morphologies
similar to that of component ``A'' are seen in galaxies where 
there has been a penetrating encounter with a smaller companion 
(e.g., the Cartwheel galaxy). Based on this, \cite{corbin05} suggest
that HS~0822+3542 is actually a system of two merging ultra-compact
dwarfs, and that it represents a dwarf galaxy in the process of 
formation. 

\begin{table}
\caption{Main parameters of the studied galaxies}
\label{tab:param}
\begin{tabular}{lcc} \\ \hline \hline
Parameter                           & HS~0822+3542            & SAO~0822+3545    \\ \hline
R.A.(J2000.0)(opt)                  & ~~08 25 55.47           & ~~08 26 05.59     \\
DEC.(J2000.0)(opt)                  & $+$35 32 32.9           & $+$35 35 25.7     \\
B$_{\rm tot}$                       & 17.92$\pm$0.05$^{(1)}$  & 17.56$\pm$0.03$^{(3)}$    \\
V$_{\rm hel}$(\ion{H}{i})(\kms)     & 726.6$\pm$2$^{(2)}$     & 740.7$\pm$1.3$^{(2)}$  \\
Dist$^{(3)}$ (Mpc)                  & 11.0                    & 11.0            \\
M$_{\rm B}^0$ $^{(4)}$              &  --12.49                & --12.85          \\
Opt. size (\arcsec)$^{(5)}$         & 14.8$\times$7.4$^{(1)}$ & 28.2$\times$15.5$^{(3)}$  \\
Opt. size (kpc)                     & 0.79$\times$0.39        & 1.50$\times$0.83$^{(3)}$  \\
Inclination (deg)$^{(6)}$           & 62                      & 59               \\
12+$\log$(O/H)  \                   & 7.44$^{(3)}$          & ---               \\
\ion{H}{i} int.flux$^{(7)}$         & 0.27$\pm$0.06$^{(2)}$   & 0.83$\pm$0.08$^{(2)}$      \\
W$_\mathrm{50}$ (km s$^{-1}$)       & 21.7$\pm$4$^{(2)}$      & 35.5$\pm$3$^{(2)}$          \\
V$_\mathrm{rot}$ (\ion{H}{i})(\kms) & $\le$5$^{(2)}$         & ~14$^{(2)}$          \\
M(\ion{H}{i}) (10$^{8} M_{\odot}$)  & 0.77$^{(2)}$            &2.36$^{(2)}$       \\
M$_{\rm dyn}$ (10$^{8} M_{\odot}$)  & 4$^{(2)}$             & 14$^{(2)}$       \\
M(\ion{H}{i})/L$_{\rm B}$$^{(8)}$   & 0.50$^{(2)}$            & 1.10$^{(2)}$       \\
\hline\hline

\multicolumn{3}{l}{(1) -- from \cite{kniazev00}} \\
\multicolumn{3}{l}{(2) -- derived in this paper; velocities are from a gaussian fit } \\
\multicolumn{3}{l}{(3) -- from \cite{pustilnik03}} \\
\multicolumn{3}{l}{(4) -- corrected for Galactic extinction A$_{\rm B}$=0.20} \\
\multicolumn{3}{l}{(5) -- $a \times b$ at $\mu_{\rm B}=$25\fm0~arcsec$^{-2}$} \\
\multicolumn{3}{l}{(6) -- from the optical diameters and assuming an intrinsic}\\
\multicolumn{3}{l}{\hskip 0.3in axis ratio of 0.2} \\
\multicolumn{3}{l}{(7) -- in units of Jy$\cdot$\kms}\\
\multicolumn{3}{l}{(8) -- in units of ($M/L_\mathrm{B}$)$_{\odot}$}  \\
\end{tabular}
\end{table}

       Our \ion{H}{i} observations show that the gas in this system 
appears to have settled into a disk with fairly regular kinematics.
We note again however, that this may be a consequence of our marginal
angular resolution. In any case, a regular HI disk may be consistent 
with the merger scenario -- recent numerical simulations 
show that in gas-rich mergers, the gas quickly settles into a disk 
(\citealp{springel05}). On the observational front, there are
cases of merging galaxies where a large fraction of the gas does seem 
to have settled into a  disk (e.g., IC~2554, \cite{koribalski03};
NGC~3310 \citealp{kregel01}). HI observations of the Cartwheel galaxy 
itself (\citealp{higdon96}) show that the gas is in a disk, albeit with
most of the gas in an expanding ring at the periphery of the disk.
While the high resolution image of HS~0822+3542 (Fig.~\ref{fig:m0ovhr}) 
could be interpreted as emission from a ring, such an intepretation
would require that the HI disk should be viewed almost perfectly
edge on, i.e. the HI disk inclination must be much higher than that inferred
from the optical image (see ~Tab.\ref{tab:param}). On the other hand,
the size scale of  the gas disk is much larger than that 
of the optical components seen in the HST image. The larger 
component~A has an angular size of only $\sim$1.6$^{''}$;
from Fig.~\ref{fig:m0ovhr}[A]) this is $\sim$26 times less than the 
deconvolved size (at a column density of 1 M$_\odot$/pc$^{2}$) 
of the \ion{H}{i} disk.  The ratio of \ion{H}{i} mass to the optical
luminosity in these
two components is however not so extreme, with 
M(HI)/(L$_{\rm B}^A + L_{\rm B}^B) \sim$1.7 (where we use the values 
for L$_{\rm B}^A + L_{\rm B}^B$ estimated by \citealp{corbin05}).
For the system as a whole we have (from Tab.~\ref{tab:param}),
M(HI)/L$_{\rm B} \sim$0.50. Note that the L$_B$ value
listed in Tab.~\ref{tab:param}  has (as is generally the case for
BCGs) a substantial contribution from nebular emission (including
line emission) while the blue  luminosities given by \cite{corbin05} 
for the two components are corrected for the contribution from
strong lines.

SAO~0822+3545 has a radial velocity difference of only $\sim$16~\kms\ 
and a projected separation of only $\sim$11~kpc from HS~0822+3542. 
The charcteristic dynamical time for this pair 
$\tau_{\rm dyn}\sim r_p/\Delta V \sim$700~Myr, comparable to the 
rotation periods of the galaxies in the pair. It would hence seem 
unreasonable to treat HS~0822+3542 as an isolated system.
The peculiar morphology of HS~0822+3542, with its large scale \ion{H}{i}
tails and the offset between the optical and \ion{H}{i} emission could
be due to interaction with  SAO~0822+3545.
The \ion{H}{i} morphology of SAO~0822+3545, as seen on the
high-resolution map in Fig.~\ref{fig:m0ovhr}[D], is also very irregular,
with a tail like feature to the north of the main body and an \ion{H}{i}
`hole' the SE part of the optical body.

   In view of the strong possibility of an external tidal trigger to
the star formation in HS~0822+3542, it is worth taking a relook 
the scenario  (earlier suggested by \citealp{pustilnik03}) where 
components ``A'' and ``B'' are star forming regions in a tidally
disturbed galaxy.  In this scenario, the ``ocular'' appearance of
component ``A'' could be the result of triggered star formation
(\cite{elmegreen77};\citealp{elmegreen02}), where feedback from  
an older central star cluster has resulted in star formation in 
a surrounding ring. In this context, it is interesting to note 
that \cite{pustilnik03} found evidence for an expanding 
superbubble in HS~0822+3542. Our \ion{H}{i} data have too 
low a signal-to-noise ratio to search for corresponding signatures in
the \ion{H}{i}  maps. However, we note that at in the highest 
resolution image (Fig.~\ref{fig:m0ovhr}[C]) the \ion{H}{i} column
density peaks on either side of the optical emission. 
Triggered star forming rings around an existing star 
cluster have been observed both in our own galaxy (\cite{deharveng03},
\citealp{deharveng05}) as well as in external dwarf galaxies 
(Sextans A; \citealp{vandyk98}).  Further, the evidence presented
by \cite{pustilnik03} for recent star formation elswhere 
in HS~0822+3542 (e.g. the arc at the  northwestern edge (or 
the tip of the ``tidal plume'' in the terminology of \citealp{corbin05})) 
is also probably easier to understand in this scenario than 
in one where HS~0822+3542 is an ongoing merger of two 
ultra-compact dwarfs. Finally, we note that though \cite{corbin05} 
state that the surface brightness of components ``A'' and ``B''
are a factor of $\sim 100$ higher than that of the stellar 
``tidal'' plume, the values for the surface brightness given 
in their paper  (23 mag arcsec$^{-2}$  for the plume and 
20.6 mag arcsec$^{-2}$ for component ``B'' and the ring in 
component ``A'') imply that the difference in surface brightness 
is only $\sim 9$. There hence appear to be good reasons
for regarding the optical knots in HS~0822+3542 as star forming
regions in a single galaxy, as opposed to merging ultra-compact
dwarfs. Higher quality data and detailed numerical modelling would
however be required to conclusively distinguish between these 
two models.

As noted by \cite{pustilnik03}, the LSB galaxy SAO~0822+3545
is unusually blue. 
Further, despite a small downward correction of its \ion{H}{i} flux
relative to that given in \cite{pustilnik03} SAO~0822+3545
remains one of the most gas-rich known LSB galaxies, with the gas-mass
fraction M$_{\rm gas}$/(M$_{\rm gas}$+M$_{\rm star}$)
in the range of 0.77 to 0.88, depending on the age/mass of the old stellar
population.  Comparison of several evolutionary models for SAO~0822+3545
with the constraints provided by its integrated colours, EW(H$\alpha$),
M$_{\rm gas}$ and M$_{\rm dyn}$ favours the scenario where the optical
emission comes from a mixture of a young stellar population 
($t_{1}$ = 10~Myr, M$_{\rm young}$ = (2.4--4.4)$\times$10$^5$ M\sunn) 
and an `old' stellar population (with age in the range of 0.25 to 
10 Gyr and mass M$_{\rm old}$ of 10-30 M$_{\rm young}$). Such a 
`starburst' event with the instantaneous formation of 
$\sim$3$\times$10$^5$M\sunn\ and a release of a large amount of
kinetic energy in the form of stellar winds and SNR should have a measurable
effect on the neutral gas; in particular one might expect to see
\ion{H}{i} ``holes''. Indeed, in Fig.~\ref{fig:m0ovhr}[D]  we see that 
in the optically brighter south-east edge there is a clear deficit of 
\ion{H}{i} gas. We note however that this deficit is seen mainly
in the high resolution image, and that it is possible that the ``hole''
is filled with low column density gas. If we do treat the feature
as a ``hole'', it corresponds to a linear size of $\sim$500~pc.
\cite{mccray87} give a relation between the hole size $R$ and 
the number of supernovae ($N_*$) producing the hole, viz.

$$R ~\sim 97 (N_* E_{51}/n_0)^{1/5} {t_7}^{3/5}$$

where $E_{51}$ is the energy per supernovae in units of $10^{51}$ erg, 
$n_0$ is the ambient density in atoms cm$^{-3}$ and $t_7$ is the 
age in units of $10^7$ yr. If we take $E_{51},n_0$ and $t_7$ 
to be $\sim 1$,  we find that the energy output from $\sim 5$ 
supernovae would be sufficient to produce a hole this size. 
This number of supernovae is easily produced by a young
stellar population with mass $\sim 2\times 10^5$M\sunn ~(from
e.g. PEGASE2; \citealp{fioc99}). While detailed modelling 
and better HI and optical data would be essential to see if the
HI ``hole'' that we see has been produced by feed back from
the star burst, the above calculation does indicate that
a young stellar population with mass similar to that inferred by  
\cite{pustilnik03} should in principle be able to
produce a hole comparable in size to the galaxy itself.

    Finally we note  that the peak inclination corrected \ion{H}{i} 
column densities in HS~0822+3542 are comparable to the observed 
threshold for star formation in dwarf galaxies 
($\NHI \sim \times 10^{21}$ atoms~cm$^{-2}$; \citealp{skillman87}). 
While star formation is indeed ongoing in HS~0822+3542, the 
peak \ion{H}{i} column density is  offset from the sites of current 
star formation (~\ref{fig:m0ovhr}[C]). This may be a 
consequence of feedback (either in the form of ionization or 
evacuation). Similarly, the \ion{H}{i} column density 
in SAO~0822+3545 (see Fig.~\ref{fig:m0ovhr}[D]) is above the
threshold value over a large fraction of its area. 
In particular, the prominent double \ion{H}{i} blob on the east
edge of the galaxy has high density, but no traces of star formation. 
There are several other known cases of dwarf galaxies where current
star formation is  offset from the location of the peak HI column 
density (e.g. SBS~0335$-$052; \citealp{pustilnik01}), and, in a detailed 
study of the relation between star formation and HI column density 
in a sample of nearby extremely faint dwarf galaxies \cite{begum06} 
found no one 
to one correspondence between high \ion{H}{i} column density and 
ongoing star formation.

     In summary, our GMRT observation of the galaxy pair
HS~0822+3542/SAO~0822+3545 shows (i)~some evidence for tidal interaction,
(ii)~that the HI emission from HS~0822+3542 has an angular size $\sim 25$ 
times larger than that of the optical components identified in the 
HST imaging, and (iii)~that the HI properties of the galaxies are in 
general similar to those of comparably sized gas rich dwarf galaxies. 
We suggest that the optical knots in HS~0822+3542 are star forming 
regions in a tidally disturbed dwarf galaxy as opposed to being 
merging ultra-compact dwarfs.

{\bf Acknowledgements}
The observations presented in this
paper were obtained using the GMRT which is operated by the National 
Centre for Radio Astrophysics (NCRA) of the Tata Institute of Fundamental
Research (TIFR), India.
SAP acknowledges partial support from the Russian state program "Astronomy"
and from the RFBR grant No.06-02-16617.
This research has made use of the NASA/IPAC Extragalactic
Database (NED), which is operated by the Jet Propulsion Laboratory,
California Institute of Technology, under contract with the National
Aeronautics and Space Administration.
The authors acknowledge the SDSS database for the images used for this study.
Funding for the Sloan Digital Sky Survey (SDSS) has been provided by 
the Alfred P. Sloan Foundation, the Participating Institutions, the 
National Aeronautics and Space Administration, the National Science 
Foundation, the U.S. Department of Energy, the Japanese Monbukagakusho, 
and the Max Planck Society. The SDSS Web site is http://www.sdss.org/.
The SDSS is managed by the Astrophysical Research Consortium (ARC) 
for the Participating Institutions. The Participating Institutions are 
The University of Chicago, Fermilab, the Institute for Advanced Study,
the Japan Participation Group, The Johns Hopkins University, the Korean 
Scientist Group, Los Alamos National Laboratory, the Max-Planck-Institute 
for Astronomy (MPIA), the Max-Planck-Institute for Astrophysics (MPA), 
New Mexico State University, University of Pittsburgh, University of 
Portsmouth, Princeton University, the United States Naval Observatory,
and the University of Washington. We thank the referee for comments that
resulted in a substantial improvement in the quality of this paper.

\bibliographystyle{mn2e}
\bibliography{lowZ}

\label{lastpage}

\end{document}